\documentstyle[prl,aps,twocolumn,epsfig]{revtex}
\begin{document}
\twocolumn[
\hsize\textwidth\columnwidth\hsize\csname @twocolumnfalse\endcsname
\draft
\title{Quantifying Nonequilibrium Behavior with Varying Cooling Rates}
\author{Carmen J. Gagne and Marcelo Gleiser}
\address{Department of Physics and Astronomy,
Dartmouth College, Hanover, NH 03755, USA }
\date{\today}
\maketitle
\begin{abstract}
We investigate nonequilibrium behavior in (1+1)-dimensional stochastic field 
theories in the context of Ginzburg-Landau models at varying cooling rates. 
We argue that a reliable measure of the departure from thermal equilibrium 
can be obtained from the absolute value of the rate of change of the 
momentum-integrated structure function, $\Delta S_{\rm{tot}}$. We show that the peak of 
$\Delta S_{\rm{tot}}$ scales with the cooling, or quench, time-scale, $\tau_q$, 
in agreement with 
the prediction by Laguna and Zurek for the scaling of freeze-out time in 
both over and under-damped regimes. Furthermore, we show that the 
amplitude of the peak scales as $\tau_q^{-6/5}$ independent of the viscosity.
\end{abstract}
]
Due to the widespread role of symmetry breaking phase transitions in cosmology, 
condensed matter physics, and high energy physics, it 
is imperative that the nonequilibrium aspects of the dynamics 
of these phase transitions be better understood.  In particular, the  
effect of the cooling rate is still 
a widely open topic of investigation.  Since
all but the simplest problems are notoriously
difficult to handle analytically, 
most approaches assume one of two extreme limits: either one assumes a
quasi-adiabatic (infinitely slow) cooling, or
an instantaneous (infinitely fast) quench, often even in 
numerical studies. However, in reality 
all cooling occurs with a finite rate and, therefore, lies somewhere
between these two extremes. Some 
notable exceptions are Wong and Knobler's \cite{Wong:1979} and 
Binder's \cite{Binder:1977} work on quenches within the one-phase region, 
Wong and Knobler's \cite{Wong:1978} and Ruiz's \cite{Ruiz:1982} work on 
double quenches, and Onuki's \cite{Onuki:1982} work on periodic quenches. 
For excellent reviews, see \cite{Gunton:1983}, \cite{Langer:1992} and 
\cite{Bray:1994}. 
More recently, Zurek, Laguna and Zurek \cite{Laguna:1998}, and Bettencourt
{\it et al.} 
\cite{Bettencourt:2000} have investigated the 
effect different cooling rates have on defect density in classical fields, while
Bowick and Momen \cite{Bowick:1998} have examined this issue for
quantum fields. 

Although these works have done much to advance our understanding of this issue,
the effect of different cooling 
rates on the dynamics of phase transitions, as well as other 
nonequilibrium situations that commonly arise in nonlinear field theories,
requires much further study.
Even if advances in computer technology in
recent years have greatly facilitated numerical studies of
phase transitions, some aspects of nonequilibrium dynamics of nonlinear field
theories remain difficult to study numerically, as they necessarily
require large lattices and/or  
long simulations. In particular, dynamical
studies near the critical region of continuous phase transitions, or 
of nucleation in discontinuous phase transitions,
require both long runs and large lattices. 

Once we focus on dynamical issues, it is often desirable to observe how
a system evolves towards its final equilibrium state. It is thus
important
to develop tools designed to 
quantify the nonequilibrium behavior of a system being cooled (or warmed up)
in a way that is numerically efficient. With this goal in mind, in this
letter we propose a possible measure to quantify the departure from
equilibrium of a system coupled to a heat bath, which clearly correlates
the approach to equilibrium with the relevant control parameters of the system,
namely, the absolute value of the rate of 
change of the momentum-integrated structure function, $\Delta S_{\rm{tot}}$,
which will be defined below.

For simplicity of comparison,
we chose to use Laguna and Zurek's model of a linear 
pressure quench for a single
classical real scalar field $\phi(x,t)$ in (1+1)-dimensions 
\cite{Laguna:1998}. 
Although there is no phase transition in one dimension, we can
still examine (local) symmetry breaking and phase ordering through the
formation of kink-antikink pairs. The present work should be considered a 
further 
step in the quantification of nonequilibrium behavior, which can be extended
to higher-dimensional systems.
All quantities have been scaled appropriately
to be dimensionless.     

We implement linear cooling with the following potential (or,
equivalently, the homogeneous 
part of the free-energy density for the order parameter $\phi$),
\begin{equation}
 V(\phi)=\frac{1}{8} \left( 1-2\epsilon(t)\phi^2 + \phi^4 \right) ~~,
\end{equation}
where $\epsilon(t)\equiv \textrm{min}(t/\tau_q,1)$, and $t$ is measured from 
the ``phase transition point'' (when $V''(\phi=0)=0$), 
so that, when the potential is a single well, both $t$ 
and $\epsilon(t)$ are negative. $\tau_q$ is the cooling time-scale. Note that 
the potential stops 
changing when $\epsilon(t)=1$, which, in a Ginzburg-Landau system, is
the zero-temperature limit.
We simulate the coupling 
of the scalar field $\phi$ to the thermal bath using a generalized 
Langevin equation,
\begin{equation}\label{e:langevin}
{\partial^2\phi\over\partial t^2} = {\partial^2\phi\over\partial x^2}- \eta
{\partial\phi\over\partial t}
- {\partial V \over \partial \phi} + \xi({\bf x},t)~~.
\end{equation}
The terminology ``pressure quench'' is justified by the fact that 
the temperature of the bath $T$, which is related to the viscosity 
$\eta$ and the stochastic force of zero mean $\xi(x,t)$ by the
fluctuation-dissipation relation,
\begin{equation}\label{e:fluct-diss}
\langle \xi(x, t) \xi(x', t') \rangle = 2 \eta T 
\delta(x - x') \delta(t - t')~,
\end{equation}
remains constant, while the quadratic 
coefficient of the potential changes linearly in time. This is equivalent to a 
pressure quench at constant temperature in the laboratory. An obvious extension 
of this work is to implement a true linear cooling, 
where the temperature of the 
bath changes linearly, and $\epsilon(t)=(T_c-T)/T_c$. We will leave this case
for a future investigation. [See, however, Ref. \cite{Bettencourt:2000}.] 
Our main interest here is in proposing a measure
for nonequilibration of classical fields, which can be adapted for several
different situations, including those involving cooling through the bath
temperature, as we show below. 

Zurek, and Laguna and Zurek have studied the effect of different cooling 
rates on the 
density of zero crossings of the field. [A counting of the number of times the
field goes through zero for a given lattice length), which provides
the approximate kink 
density as a function of time\cite{Laguna:1998}.] In general, the field will
attempt to keep up with the changing potential as best it can, that is, its
modes will try to keep thermalized as the cooling occurs. Clearly, as the
cooling rate is increased, we can
envisage a situation where this will not be possible any longer, and the
field becomes ``frozen'', unable to maintain thermal equilibrium with the bath.
According to Zurek's conjecture, this freeze-out occurs when the 
dynamical relaxation time -- given by
$\tau_{\dot \phi}\simeq |\phi/{\dot\phi}|$ for 
overdamped systems, and by $\tau_{\ddot \phi}\simeq 
|\phi/{\ddot\phi}|^{1/2}$ for 
underdamped systems -- is comparable to the time to (from) the phase transition.
From this freeze-out condition, one can derive the scaling 
relationships for the freeze-out time, 
$\hat \tau_{\dot \phi}\propto {\tau_q}^{1/3}$ 
and $\hat \tau_{\ddot \phi}\propto {\tau_q}^{1/2}$.
Using these results, Zurek and Laguna find scaling laws for kink density, which 
they confirm with simulations and contrast with experimental 
results for pressure quenches.
Here, we are mostly concerned with how
to extend our knowledge of nonequilibrium properties of field theories. We will
thus be using Zurek and Laguna's 
model as a testing ground for our methods, comparing 
some of our results to theirs.

{}For any given moment during the evolution of the system, there will be
local fluctuations around the space-averaged order parameter. These can be
studied with the structure function $S_k(t)$, which tells us how
different Fourier modes evolve in time. Its time derivative will thus give us
information on the rate of change of the individual modes.
(Notice that this is not the dynamic structure function, 
which is a Fourier transform in both space and time.) Integrating the time 
derivative of the structure function over wave number gives the 
net change of the fluctuations in the order parameter.  For simplicity, 
we use the absolute value of the 
momentum-integrated time derivative of $S_k(t)$; as we will soon see, it will
display a peak which will give us valuable information on the nonequilibrium
dynamics of the system. We thus define the quantity $\Delta S_{\rm{tot}}$ as,
\begin{equation}
\Delta S_{\rm{tot}}  = 
\left| {\int\limits_{ {2 \pi} \over L }^{ {2 \pi} \over \delta x }  
{dk\frac{{\partial S_k \left( t \right)}}{{\partial t}}} }
\right|~{\rm .}
\end{equation}
$S_k \left( t \right) = \left| {u_k \left( t \right)} \right|^2 $, where 
$u_k \left( t \right)$ is the Fourier transform of the field fluctuations, 
$u\left( {x,t} \right) = 
\phi \left( {x,t} \right) - \bar \phi \left( t \right)$ 
and $\bar \phi \left( t \right)$ is the spatial average of the field. 
$\delta x$ is the lattice spacing and $L$ is the lattice size. The 
integral is over the first Brillouin zone.  Here is our main point: In 
equilibrium the field is in a steady state, and so 
$\Delta S_{\rm{tot}} \approx 0$. Thus, values of 
$\Delta S_{\rm{tot}} $ greater than zero can be used as a measure of
nonequilibration of the system.   
$\Delta S_{\rm{tot}}$ may be smoothed, if necessary, by sampling every few 
time steps, which is equivalent to averaging over the same number of time 
steps. 

\begin{figure}
\epsfig{file=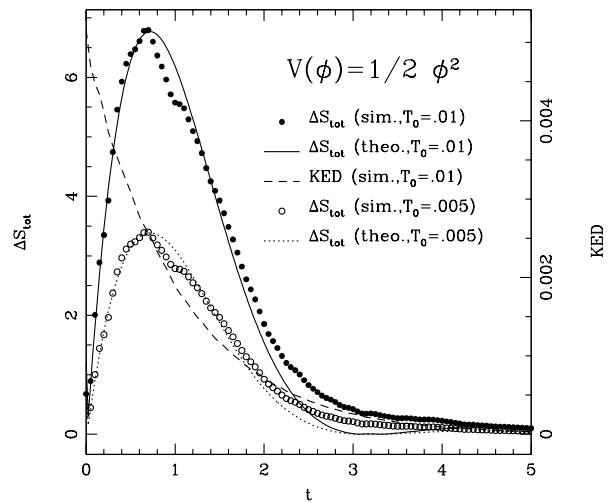,width=3.25in}
\caption{$\Delta S_{\rm{tot}}$ 
and the average kinetic energy for an instantaneous 
temperature quench
in a time-independent potential.}
\label{linplot}
\end{figure}

As a first illustration, Figure \ref{linplot} shows $\Delta S_{\rm{tot}} $ and 
the kinetic energy density, $KED$, for a field 
in a time-independent potential, $V(\phi ) = \frac{1}{2} \phi ^2 $. 
$\phi$ is initially thermalized to 
a temperature $T_0$ with viscosity $\eta$, which
is equal to unity for both cases shown.  
At $t=0$, the bath is 
instantaneously quenched to zero temperature.  Notice how initially 
$\Delta S_{\rm{tot}} $ is approximately zero  because the field is thermalized 
($KED=T_0/2$), but begins immediately to rise when the quench begins, reaches a 
maximum 
and returns to zero after the field thermalizes ($KED\rightarrow 0$).  
This case can be solved analytically 
for $\eta \le 2$, if we take the ideal limits
$\delta x \rightarrow 0 {~\rm{and}~} L \rightarrow \infty$,
\begin{eqnarray}\label{e:DSlin}
\Delta S_{tot} & \simeq  & {{{\rm{sign}}(\omega _0 t)\pi N T_0 Le^{-\eta t}}\over2}
\left|{{(1 - D^2)} \over 2}J_0 (2\omega _0 t) \right. \nonumber \\
& & \mbox{} \left. + \omega _0 D
\left[2 + D \eta - {D \over t}\right]J_1 (2\omega _0 t) +
2\omega _0 ^2 D^2 J_2 (2\omega _0 t) \right. \nonumber \\
& & \mbox{} \left. + {{\eta t} \over 2}\left( {1 + D}
\right)^2 _1 F_2 \left[{1 \over 2};1,{3 \over 2};-\omega _0^2 t^2 \right]
\right|~~,
\end{eqnarray}
where $\omega_0=\sqrt{1-{\eta ^2 \over 4}}$, $J_0$,$J_1$ and $J_2$ are 
Bessel functions of the first kind and ${}_1F_2$ is a hypergeometric function. 
$N$ and $D\equiv {C \over {4 T_0}}$, where $C$ is the value of 
$\frac{{\partial S_k \left( t \right)}}{{\partial t}}$ at $t=0$, 
are parameters that were varied to fit the data.   This is shown for both 
initial temperatures, $T_0=0.01$ and $T_0=0.005$, as a line in Figure \ref{linplot}.  
The simulation data
were fit using $N=0.82$ and $D=0.65$ for both $T_0=0.01$ and $T_0=0.005$.
We have verified (not shown) that both the time,
$t_{\rm{peak}}$, and amplitude of the peak, $A_{\rm{peak}}$,
depend on the viscosity. Also, and this is very important, the amplitude 
increases linearly with temperature change, as can be seen in Figure \ref{linplot}; 
thus, the location of the peak
gives a measure of the equilibration time-scale of the system, while its
amplitude provides a measure of the departure from equilibrium. 

\begin{figure}
\epsfig{file=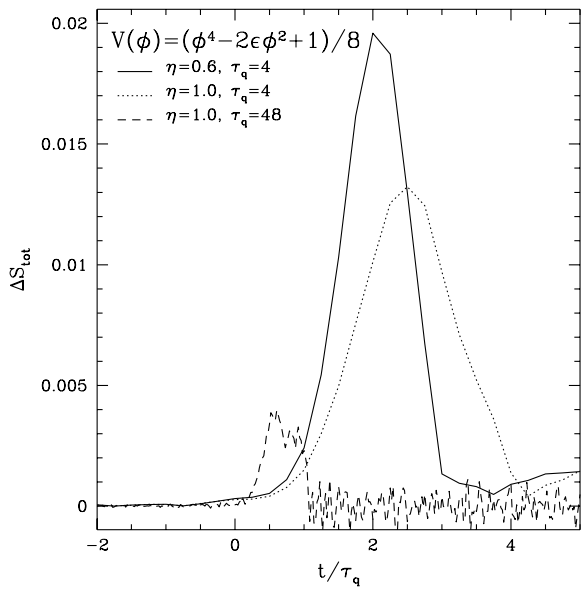,width=3.25in}
\caption{$\Delta S_{\rm{tot}}$ for different 
viscosities and cooling time-scales with the potential in Eq. 1.
$A_{\rm{peak}}$ increases with increasing cooling time-scales
and decreases with increasing 
viscosity.  $t_{\rm{peak}}/\tau_q$ 
increases with decreasing cooling time-scale and viscosity.}
\label{nlinDS}
\end{figure}

In addition to testing our second order staggered leapfrog Langevin code for 
stability 
with the same parameters as 
\cite{Laguna:1998} ($L=2048, \delta x=.125, 
\delta t=.025 \textrm{~and~}T=0.01$), 
we reproduced their defect density results to test our 
code.  For a more thorough description of the algorithm 
used see \cite{Gagne:2000}. We added the calculation of the structure 
function to our code using a FFT 
routine from Numerical Recipes \cite{Press:1992} 
and the change in the structure 
function using a simple finite difference method, 
with the ability to average over a few 
time steps to smooth the data.  We can then extract the amplitude 
and time (after the phase transition) of the peak.  Figure \ref{nlinDS} 
shows the variation in 
$\Delta S_{\rm{tot}} (t/\tau_{q } )$ with viscosity and cooling time-scale.  
$t/\tau_{q }$  is used on the time 
axis for easier comparison between different cooling time-scales.  
The amplitude of the peak is higher the
lower the viscosity and the shorter the cooling time-scale (faster cooling).
Also, the peak occurs earlier for longer cooling time-scales (slower cooling).  
Those peaks that occur during the cooling show a plateau beginning
after the peak and lasting until the cooling ends at $\epsilon=1$. We were
able to identify a peak up to $\tau_q \sim 128$ at which point the 
data becomes too noisy to distinguish the peak from the plateau. 
The fastest coolings 
(with $\eta=1$ it happens for $\tau_q \lesssim 16$ ) actually 
peak {\it after}
the potential stops changing ($\epsilon=t/\tau_q=1$): the system
remains out of equilibrium during the whole cooling process, which is
thus equivalent to an instantaneous quench.  For $\tau_q \gtrsim 256$, 
$\Delta S_{\rm tot}\simeq 0$, 
and thus the system remains thermalized during the 
cooling; this is equivalent to an adiabatic cooling. It is between these two
regimes that the cooling most dramatically affects the dynamics of the 
system.  

\begin{figure}
\epsfig{file=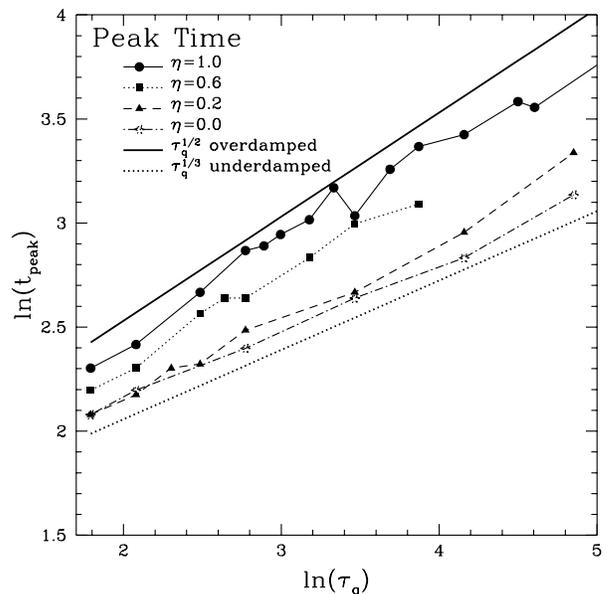,width=3.25in}
\caption{Peak time vs. cooling time-scale for time-dependent potential with
varying viscosity. The
straight lines above and
below the data show the scaling predicted in Ref. [9]
for over- and 
under-damped freeze-out times as a function of cooling time-scale. 
The simulation data interpolates between these two extremes appropriately.} 
\label{tpeak}
\end{figure}

\begin{figure}
\epsfig{file=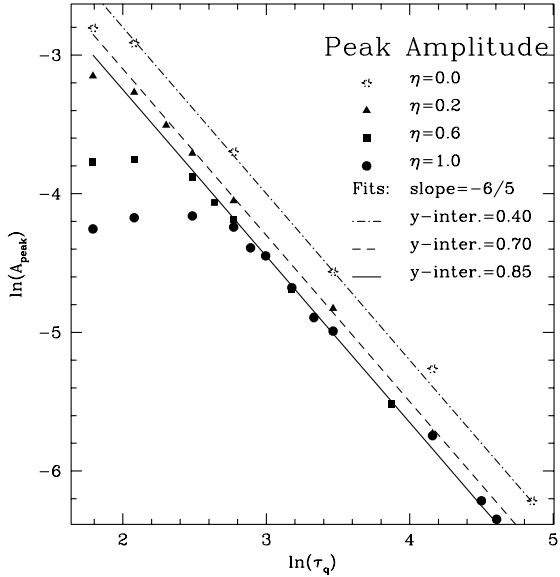,width=3.25in}
\caption{Amplitude of peak vs. cooling time-scale
for time-dependent potential with
varying viscosity. For
all but the fastest quench times, all viscosities studied satisfy the 
scaling $A_{\rm{peak}} 
\propto \tau_q^{-6/5}$. All three fit lines have a slope of -6/5; 
the fit lines are
labeled by their y-intercepts.}
\label{amp}
\end{figure}

The time and amplitude of the maximum in $\Delta S_{\rm{tot}} $ are 
plotted in Figure \ref{tpeak} and Figure \ref{amp} respectively
as a function of $\tau _q$ for several viscosities.  From fig. 3 we note that the location of the peak
scales with $\tau _q$ in agreeement to the scaling
obtained in Ref. \cite{Laguna:1998}
for the freeze-out time. 
Since this is the time after which the field is able 
to begin relaxing to its equilibrium 
state, this seems reasonable.  

We found that the amplitude of the maximum universally 
scales as $\tau _q ^{-p}$, where $p=1.2 \pm .05$ 
for all but the fastest coolings.  
There are thus essentially three regimes: i) the field remains
thermalized all through the cooling process, thus approximating an
``adiabatic'' cooling (for $\tau _q \gtrsim 256$ with $\eta=1$); 
ii) the field remains out 
of equilibrium through the complete cooling process, thus approximating
an instantaneous quench (for $\tau _q \lesssim 16$ and $\eta=1$); iii) 
the intermediate regime, where the field reaches thermalization during the 
cooling process, which is signaled by the appearance of 
a plateau in $\Delta S_{\rm{tot}}$, which decreases in amplitude as $\tau_q$
is increased, until it cannot be differentiated from noise.
The general picture can be described qualitatively 
as follows: 
the modes are initially thermalized, and their rate of change increases
as they try to ``catch up''
with the changing potential (or, when appropriate,
the changing environment).  This may or may not happen during the cooling,
determined by the value of $\tau_q$. As the quench 
proceeds at the  steady rate of $\tau_q^{-1}$, $\Delta S_{\rm{tot}}$
increases until the field begins to probe the bottom of the potential 
(the free-energy minima), when it has 
essentially ``caught up'' with the changing potential. If this happens while
the potential is still changing,  
the field then changes at the same 
rate as the potential ($\Delta S_{\rm{tot}}$ begins to plateau), until 
the potential stops changing and the field can fully thermalize 
($\Delta S_{\rm{tot}} \rightarrow 0$).  In the 
fastest coolings ($\tau_q \lesssim 16 \textrm{~for~} \eta = 1$, the field is not
able to ``catch up'' to the changing potential 
before the end of the cooling, and so $\Delta S_{\rm{tot}}$
never plateaus; the amplitude of the peak does not obey the scaling
in Figure \ref{amp} because the field is still changing after
the cooling has ended, which, as we remarked above, 
approximates an instantaneous quench. We hope to report on the extension of this
method to higher dimensional systems and on a in-depth analysis of the relevant
paramenter space in the near future.

C.G. was supported in part by a National Science Foundation Grant
no. PHY-9453431. 
M.G. was supported in part by NSF Grants PHY-0070554 and  
PHY-9453431.  C.G. thanks the Theoretical Condensed Matter Group at Boston 
University, where parts 
of this work were developed, for their hospitality and Sidney Redner, Paul
Krapivsky, Wonmuk Hwang, Prashant Sharma, Victor Spirin and Bill Klein for 
many helpful discussions.

Carmen Gagne: carmen@peterpan.dartmouth.edu

Address after 1 August 2001:  Department of Physics,

Clark University, Worcester, MA 01610, USA

Marcelo Gleiser: gleiser@dartmouth.edu

\end{document}